\documentclass[12pt,preprint,longabstract]{aastex}
\usepackage{graphicx}
\usepackage{natbib}
\usepackage{longtable} 
\newcommand{\inte}{INTEGRAL}

\newcommand{\src}{IGR~J11215$-$5952}
\newcommand{\pdot}{\.P}

\slugcomment{To appear in ApJ}

\shorttitle{SFXTs observed by {\it Swift}}
\shortauthors{L.\ Sidoli et al.}

\begin{document}
   \title{Monitoring Supergiant Fast X--ray Transients with Swift. I.  \\
Behavior outside outbursts}
    \author{L.~Sidoli\altaffilmark{1}, P.~Romano\altaffilmark{2}, V.~Mangano\altaffilmark{2}, 
A.~Pellizzoni\altaffilmark{1}, 
J.~A.~Kennea\altaffilmark{3},  
G.~Cusumano\altaffilmark{2},  
S.~Vercellone\altaffilmark{1},  
A.~Paizis\altaffilmark{1},  
D.~N.~Burrows\altaffilmark{3},  
N.~Gehrels\altaffilmark{4} 
}
 \altaffiltext{1}{INAF, Istituto di Astrofisica Spaziale e Fisica Cosmica, 
	Via E.\ Bassini 15,   I-20133 Milano,  Italy}
  \altaffiltext{2}{INAF, Istituto di Astrofisica Spaziale e Fisica Cosmica, 
	Via U.\ La Malfa 153, I-90146 Palermo, Italy} 
   \altaffiltext{3}{Department of Astronomy and Astrophysics, Pennsylvania State University, University Park, PA 16802, USA}
   \altaffiltext{4}{NASA/Goddard Space Flight Center, Greenbelt, MD 20771, USA}


\begin{abstract}
Supergiant Fast X--ray Transients (SFXTs) are a new class of High Mass X--ray Binaries (HMXBs)
discovered thanks to the monitoring of the Galactic plane performed 
with the INTEGRAL satellite in the last 5 years.
These sources display short outbursts (significantly shorter than typical Be/X-ray binaries) 
with a peak luminosity of a few 10$^{36}$~erg~s$^{-1}$. 
The quiescent level, measured only in a few sources, is around 10$^{32}$~erg~s$^{-1}$.
The X--ray spectral properties are reminiscent of those of accreting pulsars, 
thus it is likely that all the members of the new class are indeed 
HMXBs hosting a neutron star, although only 
two SFXTs have a measured pulse period, IGR~J11215--5952 ($\sim$187~s) 
and IGR~J18410$-$0535 ($\sim$4.7~s). 
Several competing mechanisms have been proposed to explain 
the shortness of these outbursts, mostly involving the
structure of the wind from the supergiant companion.
To characterize the properties of these sources on 
timescales of months (e.g.\ the quiescent level and the outburst recurrence), 
we are performing a monitoring campaign with {\it Swift}  of four SFXTs 
(IGR~J16479$-$4514,  XTE~J1739$-$302, IGR~J17544$-$2619 and AX~J1841.0$-$0536/IGR~J18410$-$0535).
{We report on the first four months of {\it Swift} observations, started on 2007 October 26.
We detect a low level X--ray activity in all four SFXTs which demonstrates that these 
transient sources accrete matter even outside their outbursts. 
This fainter X--ray activity 
is composed of many flares with a large flux variability, on timescales of thousands of seconds. 
The lightcurve variability is also evident on larger timescales of days, weeks and months, 
with a dynamic range of more than one order of magnitude in all four SFXTs. 
The X--ray spectra are typically hard, with an average 2--10 keV luminosity during this monitoring 
of about 10$^{33}$--10$^{34}$~erg~s$^{-1}$. 
We detected pulsations from the pulsar AX~J1841.0$-$0536/IGR~J18410$-$0535, with a period
of 4.7008$\pm{0.0004}$s. 
This monitoring demonstrates that these transients spend 
most of the time accreting matter, although at 
a much lower level ($\sim$100--1000 times lower) 
than during the bright outbusts, and that the ``true quiescence'', 
characterized by a soft spectrum
and a luminosity of a few 10$^{32}$~erg~s$^{-1}$,
observed in the past only in a couple of members of this class, is
probably a very rare state.
}

\end{abstract}
\keywords{X-rays: individual: (IGR~J16479$-$4514,  XTE~J1739$-$302, 
IGR~J17544$-$2619 and AX~J1841.0$-$0536/IGR~J18410$-$0535)}

\section{Introduction}

The Galactic plane monitoring performed by the INTEGRAL satellite
has led to the discovery of a number of 
new High Mass X--ray Binaries (HMXBs) in the last 5~years \citep{Bird2007}.
Several of these new sources are transients associated with OB supergiants  
and show short outbursts (a few hours, as observed with \inte, 
\citealt{Negueruela2005a}, \citealt{Sguera2006}).
These sources have been called Supergiant Fast X--ray Transients (SFXTs),
and their X--ray transient behavior is quite surprising since neutron 
stars accreting from the winds of supergiant companions were known to be persistent. 
Since their bright X--ray emission, reaching 10$^{36}$~erg~s$^{-1}$, 
is concentrated in very short outbursts, 
they are difficult to discover,
but SFXTs may be a large (and probably predominant) population of massive X--ray binaries.
Their quiescent level has been observed so far only in a few sources:
IGR~J17544$-$2619 \citep{zand2005} and IGR~J08408$-$4503  
\citep{Leyder2007}), and is at about 10$^{32}$~erg~s$^{-1}$, thus making SFXTs a class of transients
with a large dynamic range ($\sim$$10^{4}$). Among them, particularly
interesting is the case of IGR~J11215--5952, which is the only SFXT to date which displays
periodic outbursts  \citep{SidoliPM2006} with a period of 329 days (or a half of this,
as recently discovered with {\it Swift}/XRT; \citealt{Romano2007apastron}, \citealt{Sidoli2007}).
The {\it Swift} monitoring during the 2007 February outburst of \src\ 
represents the deepest and most complete set of X--ray observations of an outburst 
from an SFXT, allowed thanks to the predictable
behavior of the recurrent outbursts \citep{Romano2007}. 
These observations demonstrated that the
accretion phase during the bright outburst lasts longer than  
previously thought: a few days instead of hours, with only the brightest phase
lasting less than one day, and being characterized by a large variability with several flares
lasting from a few minutes to a few hours \citep{Romano2007}.

These observations allowed us to propose an alternative explanation 
for the SFXT outburst mechanism, 
which accounts for both the narrow shape of the IGR~J11215--5952 X--ray lightcurve and the
periodicity in the outburst recurrence. This model suggests the possible presence of
a preferential plane in the wind mass loss from the supergiant, an equatorial
wind ``disk'', which should also be inclined with respect to the orbital plane
of the binary system to explain the shortness of the outburst \citep{Sidoli2007}: 
in this framework, X--ray outbursts  are 
produced when the neutron star crosses this equatorial disk component, 
which is denser and slower than the ``polar'' wind component.
Thus, depending on the truncation of the disk, its orientation and inclination with respect
to the orbital plane, together
with the system eccentricity, the neutron star will cross once or twice the disk, resulting in 
periodic or quasi-periodic outbursts.
This implies that basically {\it all} SFXTs should display a periodicity
in the outburst recurrence, or a double periodicity, 
depending on  the different possible geometries discussed in \citet{Sidoli2007}.
We note that this scenario attempts to describe in a coherent way 
not only SFXT outbursts but also HMXBs as a class. 

Stimulated by these {\it Swift} results on \src\, 
we are performing the first sensitive wide-band X--ray 
monitoring campaign of the activity of a sample of SFXTs.
A sensitive monitoring campaign carried out as frequently as possible and 
spanning very long timescales is crucial
to fully characterize the behavior of  SFXTs, to determine the 
properties of their quiescent state (where the accumulation of
large observing time is needed to allow a meaningful spectral analysis of this faintest emission), 
to monitor the onset of the outbursts and measure the outburst recurrence period(s). 
Determination of the outburst pattern 
is fundamental in order to discover the supposed periodicities (or quasi-periodicities) 
implied by our proposed model. A positive (or negative) result will be possible only after
several months  of observations (for example, the period for the outburst recurrence 
in IGR~J11215--5952 is $\sim$165~days). This will allow us to reach a firm conclusion 
about periodic or non-periodic behavior. 
A monitoring with these aims  is now possible 
only with the {\it Swift} satellite, which  offers a good sensitivity combined with 
a broad-band coverage (from optical/UV to hard X--rays) and, most importantly, a unique flexibility.
This latter becomes crucial when observing transients with short outbursts, in order to
rapidly modify the original schedule when they show indications of 
an imminent outburst (possible ``bounces'' in the light curve, 
which would normally be missed), to catch not only the ourburst but also its onset. 

Here we report on the results of the first four months of this on-going 
campaign with {\it Swift}, which targeted the following sources: 
IGR~J16479$-$4514,  XTE~J1739$-$302, IGR~J17544$-$2619 and IGR~J18410$-$0535
(see \citealt{Walter2007} and references therein, for an updated review of the parameters 
of these systems).
The targets have been selected considering sources which,
among several candidates of this newly discovered class of sources,  are
confirmed SFXTs, i.e.\ they display both a ``short'' transient (and recurrent) 
X--ray activity 
and they have been optically identified with supergiant companions (see references in \citealt{Walter2007}).
In particular, XTE~J1739-302 and IGR~J17544$-$2619 are generally 
considered prototypical SFXTs: 
XTE~J1739$-$302 was the first transient which showed
an unusual X--ray behavior \citep{Smith1998:17391-3021}, only recently optically associated
with a blue supergiant \citep{Negueruela2006}.
IGR~J16479$-$4514 has displayed in the past a more frequent X--ray outburst occurrence than other
SFXTs,
(see, e.g. Walter et al. 2007),
and offers an {\it a priori} better chance to be caught during an outburst.   
AX~J1841.0$-$0536/IGR~J18410$-$0535, on the other hand, is an interesting source which, 
together with IGR~J11215--5952, is the only
SFXT where a pulsar has been detected \citep{Bamba2001} and which may offer the opportunity  
to determine the orbital parameters from the pulsar timing on long timescales.

The results we are reporting here are concentrated on the {\it out-of-outburst} 
behavior of the four SFXTs monitored with {\it Swift} between 2007 October 26 and 
2008 February 28.
An old outburst from IGR~J16479$-$4514 observed in 2005 
in archival {\it Swift} data is also discussed.

\section{Observations and data analysis: new and archival data}

Table~\ref{sfxts:tab:alldata} reports the log of the {\it Swift} observations,
which include both the new and the old ones, retrieved from the {\it Swift} Archive. 

The XRT data were processed with standard procedures ({\tt xrtpipeline}
v0.11.6), filtering, and screening criteria by using FTOOLS in the
{\tt Heasoft} package (v.6.4).
We considered both WT and PC data, depending on the count rate of the sources,
and selected event grades 0--2 and 0--12, respectively.
When appropriate, we corrected for pile-up.
To account for the background, we also extracted events within 
source-free regions.
Ancillary response files were generated with {\tt xrtmkarf},
and they account for different extraction regions, vignetting,  and
PSF corrections. We used the latest spectral redistribution matrices
(v010) in the Calibration Database maintained by HEASARC.
For timing analysis, the arrival times of XRT events were
converted to the Solar System barycentre with the task
{\tt barycorr}.

During the observations of our monitoring campaign BAT observed
the sources simultaneously with XRT, and survey data products, 
in the form of detector plane histograms (DPH), are available.
IGR J16479$-$4514 \citep{Kennea2005:16479-4514,Kennea2006:transients,Markwardt2006:16479-4514} 
and XTE~J1739$-$302 also triggered the BAT
in the past (see Table~\ref{sfxts:tab:alldata}, observations labelled with (a))  
and in those cases BAT events are also available.
The optical counterparts of the sources are expected 
to display a flux at a level of V$\sim$ 16--20 mag, 
and have been observed by UVOT. 
The BAT and UVOT data will be included in a forthcoming paper.

\section{Results}

\subsection{Lightcurves}

Figures~\ref{sfxts:fig:lcv_all}(a) to~\ref{sfxts:fig:lcv_all}(f) show
the results as lightcurves in the 0.2--10\,keV band,
which were corrected for pile-up, PSF losses, and vignetting and background-subtracted.
Each point in the lightcurves of Figure~\ref{sfxts:fig:lcv_all}(b)  
to ~\ref{sfxts:fig:lcv_all}(f) refers to the average flux observed during 
each snapshot observation performed with {\it Swift}/XRT and  
reported in Table~\ref{sfxts:tab:alldata}; 
each point in Figure~\ref{sfxts:fig:lcv_all}(a) 
has at least 25 source counts per bin and minimum time bin of 15\,s, 
while the insets in Figures~\ref{sfxts:fig:lcv_all}(b) and Figure~\ref{sfxts:fig:lcv_all}(e)
have at least 18 source counts per bin and minimum time bin of 30\,s.
These insets show details of specific observations. 
The downward-pointing arrows are 3-$\sigma$ upper limits. 
The lack of observations occurring roughly from December 2007 to January 2008,
depending on the target coordinates, is due to the sources being Sun-constrained.
We show both archival data as black points (triangles)  
and the ToOs we requested in the last four months 
(from 2007 October 26, until till the end of 2008 February) as red points (filled circles). 
In particular, Figure~\ref{sfxts:fig:lcv_all}(a) 
shows an intense outburst of IGR~J16479$-$4514 which triggered BAT in 2005
(image trigger, trigger number 152652); in that instance, this source
was detected with BAT for about 60\,s, with an average significance level of 8.7~$\sigma$ 
in the 15--100\,keV energy range, and an average count rate of 
(7.7$\pm{0.9})$$\times$10$^{-3}$ counts$^{-1}$~s$^{-1}$~det$^{-1}$;
unfortunately this source was observed only once during the fall 2007
monitoring campaign because of Sun constraints 
(Figure~\ref{sfxts:fig:lcv_all}(b)). 

Figures~\ref{sfxts:fig:lcv_all}(c) and \ref{sfxts:fig:lcv_all}(d) 
show the full dataset on IGR~J17544$-$2619. 
We note that an outburst was discovered in BAT 
Monitor\footnote{http://swift.gsfc.nasa.gov/docs/swift/results/transients/index.html}  
data on MJD~54412 \citep{Krimm2007:ATel1265},
merely two days after the source had become unobservable by the XRT 
because it was Sun-constrained.
Figs.~\ref{sfxts:fig:lcv_all}(e) and \ref{sfxts:fig:lcv_all}(f) show the
light curves of XTE~J1739$-$302 and IGR~J18410$-$0535, respectively.

Bamba et al. (2001) discovered coherent pulsations with a period of
4.7394$\pm$0.0008 s in the brightest flare phase of the transient X--ray
pulsar AX~J1841.0$-$0536/IGR~J18410$-$0535.
In order to search for periodicities, we considered 
the {\it Swift}/XRT observations with the best signal to noise 
(observations ID 00030988001,
00030988004 and 00030988005 in Table\ref{sfxts:tab:alldata}). 
We examined a  frequency range centered on the above value, performing
an epoch folding technique on this subset of data. 
The best-fit pulse period is 4.7008$\pm$0.0004 s. The corresponding
Pearson statistics for the pulse histogram with 10 bins ($\sim$20 mean
source counts per bin) provides a reduced $\chi^{2}$ of 6.3 (9 degrees
of freedom) which, even taking into account the $\sim$$10^3$ searched
periods, has a virtually null probability of chance occurrence.


\subsection{Spectra}

The long term lightcurves of all sources show a large variability
over time scales of hours, days and weeks, 
with a  frequent low level flaring activity. 
In order to characterize the spectral properties of the sources in this
fainter state, we extracted a single spectrum for each source, 
integrating over all the available observing time when the source is not in outburst.
The average spectral parameters are reported in Table~\ref{sfxts:tab:spectra}. 
They display hard and highly absorbed spectra (see Fig.~\ref{sfxts:fig:spec_all}), 
with absorption in excess
of the total interstellar Galactic value towards the sources. 

The spectrum of the source IGR~J16479$-$4514 during the bright outburst
reported in Fig.~\ref{sfxts:fig:lcv_all}(a) is fit well by a highly 
absorbed (column density of 4--10\,$\times$10$^{22}$~cm$^{22}$) 
hard power law (photon index, $\Gamma$, in the range 0.6--1.4). 
During the fainter state monitored in 2008 (see Fig~\ref{sfxts:fig:spec_all}(a)),
the spectrum appears softer ($\Gamma=1.4$--$1.8$) and equally highly absorbed. 
There is no evidence for a variable absorbing column density in
IGR~J16479$-$4514 between the long term low level emission and the bright outburst 
observed in 2005.

\section{Discussion}

We report here on the results of the first  
sensitive long term X--ray monitoring
of SFXT lightcurves, performed with {\it Swift}, focussing
on the out-of-outburst X--ray behavior.

The first months of these observations of a sample of
four transients reveal that they spend most of the time in a low-level
X--ray activity which is far from being a true quiescent state.
All four of these sources display, outside the bright outbursts, 
a highly variable low level  X--ray activity, 
characterized by a flaring behavior with
a large dynamic range: more than one order of magnitude for all the sources. 
The average spectra of this  X--ray emission are characterized by 
hard powerlaw photon indexes (in the range 1--2), high absorbing column densities, 
and  average fluxes which translate into X--ray luminosities of a few 10$^{33}$~erg~s$^{-1}$ 
(see Table~\ref{sfxts:tab:spectra}), assuming the
distances optically determined by \citet{Rahoui2008}), up to an average 
level of 6--8$\times$10$^{34}$~erg~s$^{-1}$ observed in IGR~J16479$-$4514.

Previous X--ray observations of SFXTs outside outbursts consist
of a few short exposures with ASCA, {\it Swift}/XRT, {\it Chandra}, XMM-{\it Newton},
which were not part of a systematic monitoring campaign.
During these observations the true quiescent state was caught
in IGR~J17544$-$2619 \citep{zand2005}, in IGR~J08408$-$4503
\citep{Leyder2007},
and probably in XTE~J1739$-$302 (\citealt{Sakano2002:ASCA}).
This quiescence is  characterized by a luminosity  of about 10$^{32}$~erg~s$^{-1}$,
and by a very soft spectrum: for example, in IGR~J17544$-$2619, the fit  with a
power law resulted in a photon index of almost 6 \citep{zand2005}.
The kind of low-level activity we observe now with {\it Swift} on long timescales of months,  
has only been observed before in single short exposures targeted on 
IGR~J17544$-$2619 \citep{Gonzalez2004} and
on  XTE~J1739$-$302 \citep{smith2006aa}, during observations
which covered only a few tens of ks. 
Our {\it Swift} monitoring campaign, after four months of 
observations (on average 1--2\,ks of net exposure per source twice or three times a week depending
on the source), has already demonstrated that accretion at a low level rate is a very
frequent and more typical state in SFXTs, and that instead the quiescence seems to be a much
rarer behavior  for these sources than previously thought.
These findings 
firmly establish that SFXTs cannot be considered as sources which 
undergo long periods of quiescence only occasionally 
interrupted by the sudden accretion of matter from the wind 
of the supergiant companion. 
Instead, they continue accreting matter
even when they are not in outburst, over long timescales of months, with a large
variability in the X--ray flux.

We measured the pulsar period in XRT data of the SFXT IGR~J18410$-$0535. 
Compared with the period determined in ASCA data by \citet{Bamba2001}, there
is evidence for a spin-up trend, with a pulse period difference of 
$\Delta$P=$-$0.0386$\pm{0.0009}$~s, and an average period derivative, \pdot,
of $-1.5\times10^{-10}$~s~s$^{-1}$, between the two determinations separated by 
$\sim$2965~days.
Since the orbital parameters are unknown, it is not possible to correct for the Doppler delay
due to the orbital motion in the binary system. Moreover, a spin-up trend
is often observed in HMXBs (see e.g. \citealt{Bildsten1997}) 
induced by the accretion torque \citep{Ghosh1979}, thus at this stage it is not
possible to discriminate between the two effects. 
To estimate the maximum possible shift on the pulse period due to the orbit,
we can assume that
the pulsar is orbiting near to the surface of the OB supergiant companion; in this
case the shift is about 0.2\% difference
between the two period measurements, which is less than what we observe, thus 
a good fraction of the pulse period change is real.
More determinations of the spin period are needed to characterize the orbit, 
and will hopefully be
available in the next months of this observing campaign. On the other hand,
besides the large intensity variability and hard spectral properties, this is
a further confirmation of the fact that this source
is still accreting matter even at much lower rates than
in outburst.

The main hypotheses proposed 
to explain the SFXTs behavior (especially their outburst durations) are
based on the structure of the wind of the OB supergiant companions.
In't Zand (2005) originally proposed that the sudden accretion of material 
from the clumpy wind of the supergiant could give rise to the bright short outbursts.
Negueruela et al. (2005) suggested that SFXTs orbits should be highly eccentric
to explain the low luminosity in quiescence. More recently, the idea of a clumpy and
spherically symmetric wind has been associated with an eccentric and/or wide orbit
(Walter \& Zurita Heras 2007, \citealt{Negueruela2007}). 
For these authors the main difference between the SFXTs and  the persistent HMXB
depends on the number of clumps encountered and accreted 
by the compact object along its orbit:
in persistent systems the rate is very high because the orbit lies within $\sim$2 stellar radii
from the supergiant, while in the SFXTs the orbit probably lies at higher distances, 
in a region where the wind clumps density is much lower. 
\citet{Negueruela2008} suggest that outside this radius the space is
effectively void, and the probability for the neutron star to accrete a wind clump very low. 
In this model the SFXT outbursts are produced by the accretion of a single clump
and the X--ray luminosity can be used 
to determine the mass of the clump, knowing the flare duration \citep{Walter2007}.
In this case, the mass of the single clumps accreted to explain the low level flaring activity
we observe with {\it Swift} should be about two orders of magnitude 
lower than during the bright outburst (where the luminosity exceeds 10$^{36}$~erg~s$^{-1}$), 
assuming the
same flare duration.

A second hypothesis has been proposed by \citet{Sidoli2007} and is based on
the shape of the lightcurve observed during the 2007 outburst from the
unique SFXT displaying periodic outbursts, \src. 
This set of observations clearly demonstrated that the SFXTs outbursts cannot be 
produced by the accretion of a single clump, nor by a spherical distribution
of clumps. 
The X--ray lightcurve of the 2007 February outburst was too narrow and steep to be explained
by an enhanced accretion rate when the neutron star approaches the periastron passage
orbiting a supergiant with a spherically symmetric wind,
even in a binary system with an extremely high eccentricity.
This strongly suggested some degree of anisotropy in the supergiant wind,
which could be explained by the presence of a second denser wind component 
(besides the polar spherically symmetric wind) 
in the form of an equatorial wind ``disk'' from the supergiant donor.
The shortness of the outburst is also indicative of an equatorial wind which is inclined
with respect to the orbital plane. A flaring activity was detected and could be explained
with a wind that is inhomogeneous in addition to being anisotropic \citep{Sidoli2007}.

Basically, the main difference between the two hypotheses resides indeed in the
spherical (e.g. \citealt{Negueruela2008}) vs.\  anisotropic wind \citep{Sidoli2007}. 
In both models the low-level X--ray activity could be explained:
in the first scenario, it  could probably be produced by clumps with a distribution of 
sizes and masses (breakup of clumps
to smaller sizes at larger distances in the eccentric case, or a distribution of
clump sizes in the non-eccentric case),
while in the anisotropic wind scenario the low level X--ray emission
is very easily explained by the fact that the
neutron star, when not undergoing a bright outburst, is not crossing 
the ``disk'' wind component from the supergiant donor, but in any case
accretes matter
from the polar supergiant wind, which is faster and two orders of magnitude less dense than
the equatorial wind component.

{\it Facilities:} \facility{{\it Swift} (XRT)}.

\acknowledgements
We thank the {\it Swift} team for making these observations possible,
in particular the duty scientists and science planners.
This work was supported by MIUR grant 2005-025417, contracts  
ASI/INAF I/023/05/0, I/008/07/0 and I/088/06/0, and {\it Swift} NASA contract NAS5-00136.
PR thanks INAF-IASFMi, where part of the work was carried out, for their kind hospitality. 
We thank the anonymous referee for a swift and very constructive report.


\setcounter{figure}{0}  
        \begin{figure*}
	\epsscale{0.8}
	\vspace{-1truecm}
        \plotone{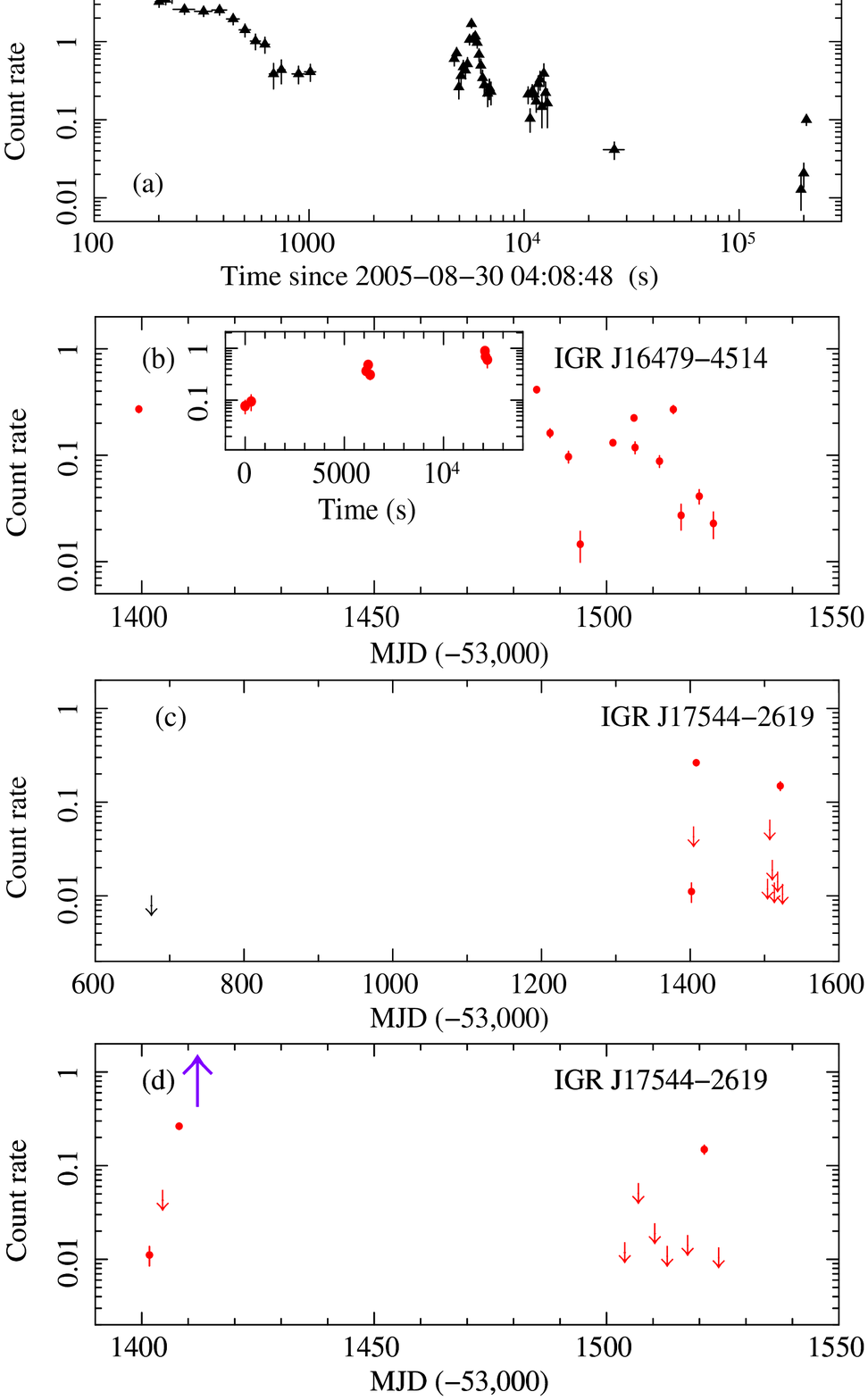}
	\caption[]{{\it Swift}/XRT (0.2--10\,keV) light curves, corrected for pile-up, 
                PSF losses, vignetting and background-subtracted. 
		The black triangles are archival data, i.e.\ observations performed before our 
		monitoring started, and red circles are 
		the data from the ToOs we requested (from 2007 October 26, until 
		the end of 2008 February). 
		The downward-pointing arrows are 3-$\sigma$ upper limits. 
		{\bf (a)} Archival data from the 2005 outburst of IGR~J16479$-$4514, referred to the
			BAT trigger (2005-08-30 04:08:48 UT). 
		{\bf (b)} Data on IGR~J16479$-$4514 collected in 2007 and 2008. 
			 The inset shows a detail of observation 014 (centered on MJD 54514.35). 
		{\bf (c)} Full data set on IGR~J17544$-$2619.
		{\bf (d)} Data on IGR~J17544$-$2619 collected in 2007 and 2008. 
		Note that an outburst was discovered in BAT Monitor data on MJD 54412 (blue upward pointing arrow). 
                      [See the electronic edition of the
                      Journal for a color version of this figure.]
		      \label{sfxts:fig:lcv_all}          }
        \end{figure*}

\clearpage 
\setcounter{figure}{0}
        \begin{figure*}
	\epsscale{0.8}
	\vspace{-5truecm}
	\plotone{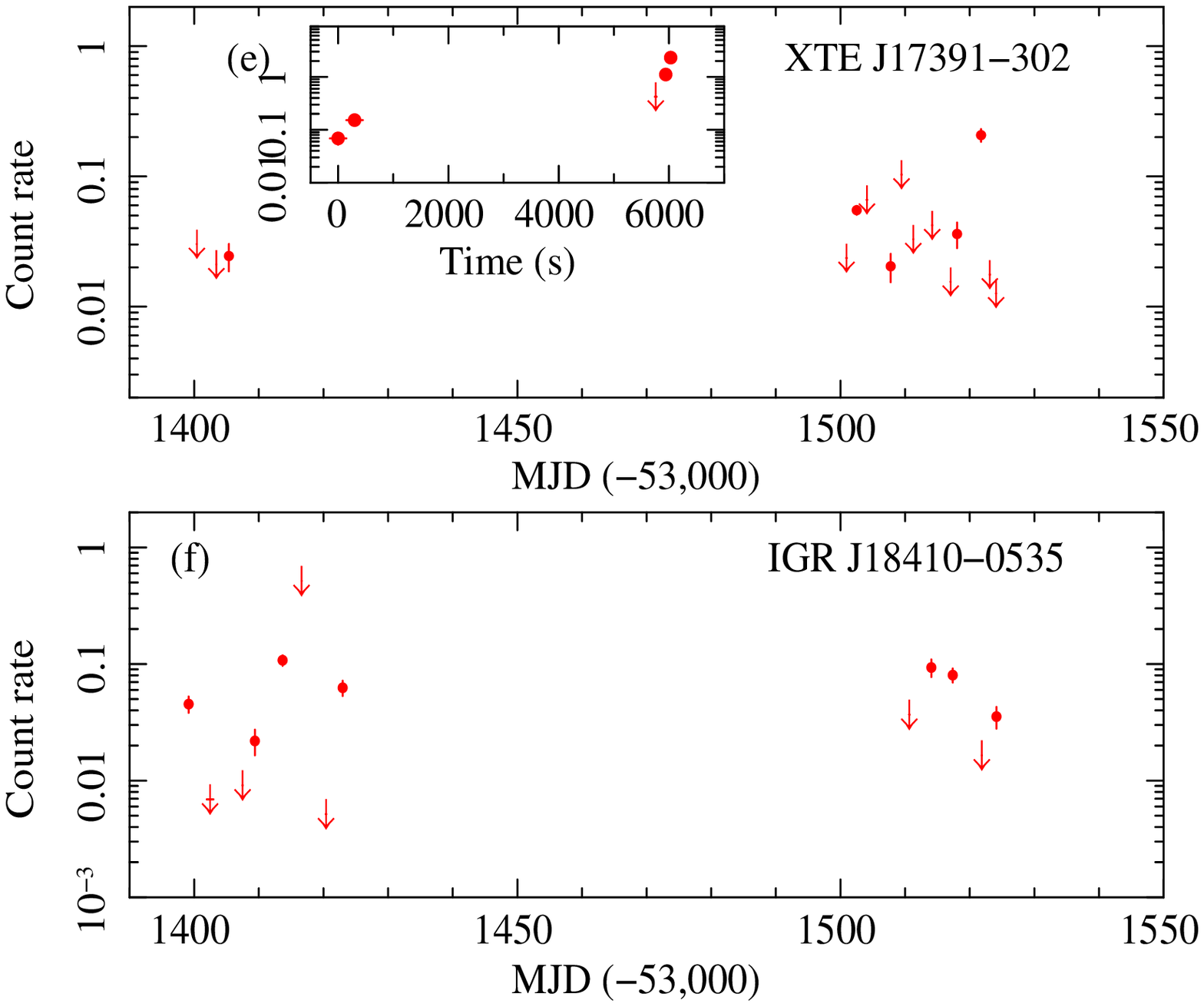}
	\vspace{-5truecm}
	\caption[]{Continued. 
		{\bf (e)} Full data set on XTE~J1739$-$302 (2007 and 2008). 
			  The inset shows a detail of observation 013 (MJD 54521.76). 
		{\bf (f)} Full data set on IGR~J18410$-$0535 (2007 and 2008). [See the electronic edition of the
                      Journal for a color version of this figure.]
		      \label{sfxts:fig:lcv_all_b}          }
        \end{figure*}

        \begin{figure*}
	\epsscale{1.0}
	\vspace{-5truecm}
	\plotone{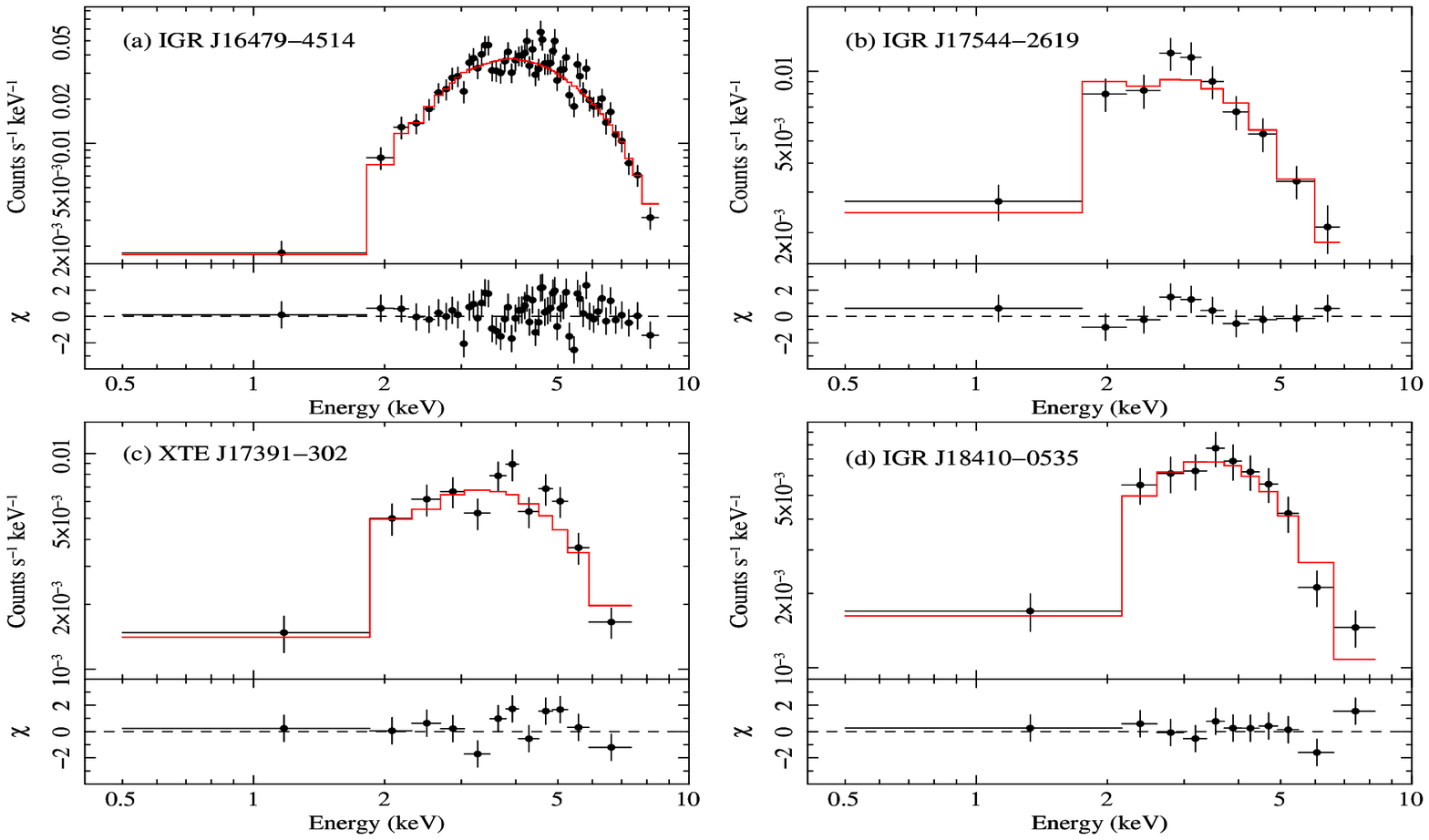}
	\vspace{-5truecm}
                \caption[]{{\it Swift}/XRT spectra.
		{\bf Top:}  out-of-outburst data from the four SFXTs in our sample, fit with an absorbed power-law model. 
		{\bf Bottom:} the residuals of the fit (in units of standard deviations).  [See the electronic edition of the
Journal for a color version of this figure.]
		\label{sfxts:fig:spec_all}   }
        \end{figure*}

\begin{deluxetable}{llllll}
  \tabletypesize{\scriptsize}
  \tablewidth{0pc} 	      	
  \tablecaption{Observation log.\label{sfxts:tab:alldata}} 
  \tablehead{
\colhead{Name} & \colhead{Sequence\tablenotemark{a}} & \colhead{Instrument/Mode} & \colhead{Start time (UT)} &  \colhead{End time (UT)} &  
               \colhead{Net Exposure\tablenotemark{b}} \\
\colhead{}       & \colhead{}            & \colhead{}    & \colhead{(yyyy-mm-dd hh:mm:ss)} & \colhead{(yyyy-mm-dd hh:mm:ss)} &  \colhead{(s)} \\
\colhead{(1)}    & \colhead{(2)}         & \colhead{(3)} & \colhead{(4)}         & \colhead{(5)} & \colhead{(6)}
}
  \startdata
IGR~J16479$-$4514& 00152652000\tablenotemark{c}	&BAT/evt &2005-08-30 04:03:49	&2005-08-30 04:13:51 	& 	602 	\\
		& 00030296001			&XRT/WT	 &2005-08-30 04:11:00	&2005-09-01 13:40:51	&	4541	\\
		& 00030296001			&XRT/PC	 &2005-08-30 04:12:41	&2005-09-01 14:01:58	&	11408	\\
		& 00030296002			&XRT/PC	 &2005-09-10 00:15:00	&2005-09-12 12:10:59	&	6394	\\
		& 00030296003			&XRT/PC	 &2005-09-14 00:43:34	&2005-09-14 10:47:57	&	4182	\\
		& 00030296004			&XRT/PC	 &2005-10-18 09:17:15	&2005-10-18 14:35:57	&	5423	\\
		& 00210886000\tablenotemark{c}   	&BAT/evt &2006-05-20 17:32:29   &2006-05-20 17:33:12    &       43	\\
		& 00215914000\tablenotemark{c}  	&BAT/evt &2006-06-24 20:19:49   &2006-06-24 20:20:32    &       43	\\
		& 00286412000\tablenotemark{c}   	&BAT/evt &2007-07-29 12:07:25   &2007-07-29 12:08:08    &       43	\\
		& {\bf 00030296005}             &XRT/PC &2007-10-26 08:08:36	&2007-10-26 09:42:57	&	1176	\\
		& {\bf 00030296006}     	&XRT/PC  &2008-01-19 22:58:28   &2008-01-19 23:13:58    &       927     \\
		& {\bf 00030296007}     	&XRT/PC  &2008-01-22 20:07:57   &2008-01-22 20:25:58    &       1079    \\
		& {\bf 00030296008}     	&XRT/PC  &2008-01-26 18:56:04   &2008-01-26 19:14:57    &       1133    \\
		& {\bf 00030296009}     	&XRT/PC  &2008-01-29 08:03:13   &2008-01-29 08:20:56    &       1062    \\
		& {\bf 00030296010}     	&XRT/PC  &2008-02-05 00:48:59   &2008-02-05 16:58:56    &       3056  \\ 
		& {\bf 00030296011}	&XRT/PC	&	2008-02-09 18:53:39	&	2008-02-09 23:50:58	&	2749	\\
		& {\bf 00030296012}	&XRT/PC	&	2008-02-10 01:16:40	&	2008-02-10 04:36:58	&	873	\\
		& {\bf 00030296013}	&XRT/PC	&	2008-02-15 08:12:38	&	2008-02-15 10:01:25	&	1254	\\
		& {\bf 00030296014}	&XRT/PC	&	2008-02-18 06:49:05	&	2008-02-18 10:12:57	&	804	\\
		& {\bf 00030296014}	&XRT/WT	&	2008-02-18 06:48:56	&	2008-02-18 10:10:23	&	37	\\
		& {\bf 00030296015}	&XRT/PC	&	2008-02-20 00:33:38	&	2008-02-20 00:46:58	&	798	\\
		& {\bf 00030296016}	&XRT/PC	&	2008-02-23 21:44:31	&	2008-02-23 23:35:57	&	1667	\\
		& {\bf 00030296017}	&XRT/PC	&	2008-02-26 23:38:06	&	2008-02-26 23:52:56	&	890	\\
IGR~J17544$-$2619 &00035056001				&XRT/PC	&2005-11-01 01:26:13	&2005-11-01 23:59:59	&	2633	\\
		&{\bf 00035056002}			&XRT/PC	&2007-10-28 00:20:09	&2007-10-29 07:07:56	&	2783	\\
		&{\bf 00035056003}			&XRT/PC	&2007-10-31 10:19:05	&2007-10-31 13:43:35	&	248	\\
		&{\bf 00035056004}			&XRT/PC	&2007-11-04 00:58:32	&2007-11-04 01:16:58	&	1104	\\
		& \tablenotemark{d}			&BAT/MON &2007-11-08 01:31:04	&2007-11-08 06:07:11	& 	\nodata	\\
		&{\bf 00035056005}			&XRT/PC &2008-02-07 20:17:58	&2008-02-07 23:43:57	&	1331	\\
		&{\bf 00035056006}	&XRT/PC	&2008-02-10 19:14:07	&	2008-02-10 20:52:56	&	278	\\
		&{\bf 00035056007}	&XRT/PC	&2008-02-14 08:17:17	&	2008-02-14 10:00:56	&	915	\\
		&{\bf 00035056008}	&XRT/PC	&2008-02-17 00:16:44	&	2008-02-17 02:04:56	&	1346	\\
		&{\bf 00035056009}	&XRT/PC	&	2008-02-21 10:24:08	&	2008-02-21 10:40:36	&	960	\\
		&{\bf 00035056010}	&XRT/PC	&	2008-02-24 23:41:09	&	2008-02-25 01:24:57	&	998	\\
		&{\bf 00035056011}	&XRT/PC	&	2008-02-28 02:41:54	&	2008-02-28 02:58:57	&	1004	\\
XTE~J1739$-$302 &00282535000\tablenotemark{c} &BAT/evt  &2007-06-18 03:10:37	&2007-06-18 03:11:20	&	43	\\
		&{\bf 00030987001}		&XRT/PC	&2007-10-27 09:46:16	&2007-10-27 10:05:57	&	1181	\\
		&{\bf 00030987002}		&XRT/PC	&2007-10-30 10:12:32	&2007-10-30 10:25:57	&	805	\\
		&{\bf 00030987003}		&XRT/PC	&2007-11-01 08:47:14	&2007-11-01 09:06:58	&	1181	\\
		&{\bf 00030987004}     		&XRT/PC &2008-02-04 21:24:07    &2008-02-04 22:57:58    &       940     \\
		&{\bf 00030987005}		&XRT/PC &2008-02-06 00:48:04	&2008-02-06 23:29:58	&	5606	\\
		&{\bf 00030987006}	&XRT/PC	&	2008-02-08 01:10:03	&	2008-02-08 02:48:58	&	161	\\
		&{\bf 00030987006}	&XRT/WT	&	2008-02-08 01:09:57	&	2008-02-08 02:46:28	&	11	\\
		&{\bf 00030987007}	&XRT/PC	&	2008-02-11 17:26:01	&	2008-02-11 19:16:56	&	1295	\\
		&{\bf 00030987008}	&XRT/PC	&	2008-02-13 09:48:13	&	2008-02-13 11:30:54	&	180	\\
		&{\bf 00030987009}	&XRT/PC	&	2008-02-15 06:32:34	&	2008-02-15 06:40:58	&	504	\\
		&{\bf 00030987010}	&XRT/PC	&	2008-02-18 03:45:06	&	2008-02-18 05:26:56	&	498	\\
		&{\bf 00030987011}	&XRT/PC	&	2008-02-21 00:40:01	&	2008-02-21 00:57:57	&	1076	\\
		&{\bf 00030987012}	&XRT/PC	&	2008-02-22 00:52:08	&	2008-02-22 01:06:55	&	888	\\
		&{\bf 00030987013}	&XRT/PC	&	2008-02-25 17:19:52	&	2008-02-25 19:03:57	&	597	\\
		&{\bf 00030987014}	&XRT/PC	&	2008-02-27 01:26:01	&	2008-02-27 03:05:57	&	937	\\
		&{\bf 00030987015}	&XRT/PC	&	2008-02-28 01:23:19	&	2008-02-28 01:40:57	&	1056	\\
IGR~J18410$-$0535 &{\bf 00030988001}	&XRT/PC	&2007-10-26 00:08:53	&2007-10-26 06:45:56	&	1384	\\
		&{\bf 00030988002}	&XRT/PC	&2007-10-28 22:53:11	&2007-10-29 23:12:56	&	3199	\\
		&{\bf 00030988003}	&XRT/PC	&2007-11-03 10:37:52	&2007-11-03 12:24:56	&	1122	\\
		&{\bf 00030988004}	&XRT/PC	&2007-11-05 09:08:39	&2007-11-05 09:28:58	&	1218	\\
		&{\bf 00030988005}     	&XRT/PC &2007-11-09 16:11:30    &2007-11-09 16:32:56    &       1286  \\
		&{\bf 00030988006}	&XRT/PC	&2007-11-12 14:51:25	&2007-11-12 14:51:55	&	30	\\
		&{\bf 00030988007}	&XRT/PC	&2007-11-16 08:40:04	&2007-11-16 10:36:58	&	2332	\\
		&{\bf 00030988008}	&XRT/PC	&2007-11-18 23:22:32	&2007-11-18 23:41:58	&	1165	\\
		&{\bf 00030988009}	&XRT/PC	&	2008-02-14 14:44:51	&	2008-02-14 16:28:58	&	947	\\
		&{\bf 00030988009}	&XRT/WT	&	2008-02-14 14:42:45	&	2008-02-14 16:20:15	&	127	\\
		&{\bf 00030988010}	&XRT/PC	&	2008-02-18 00:26:32	&	2008-02-18 02:13:57	&	626	\\
		&{\bf 00030988011}	&XRT/PC	&	2008-02-21 08:46:13	&	2008-02-21 09:02:58	&	1005	\\
		&{\bf 00030988012}	&XRT/PC	&	2008-02-25 20:32:34	&	2008-02-25 20:47:57	&	834	\\
		&{\bf 00030988013}	&XRT/PC	&	2008-02-28 03:00:08	&	2008-02-28 03:16:56	&	1008
  \enddata 
  \tablenotetext{a}{Bold-faced observations are new; the others are archival data. }
  \tablenotetext{b}{The exposure time is spread over several snapshots  
	(single continuous pointings at the target) during each observation.}
  \tablenotetext{c}{The source triggered {\it Swift}/BAT. }
  \tablenotetext{d}{The source was detected with the Swift/BAT hard X-ray transient monitor.}
  \end{deluxetable}


\begin{table*}
\caption{XRT spectroscopy of the four SFXTs, out of outburst (2007+2008 dataset). 
The data are fit  with two models:
an absorbed power law  and an absorbed blackbody.
$\Gamma$ is the power-law photon index, $kT_{\rm bb}$ is the blackbody temperature. 
The column $N_{\rm H}$ is in units of 10$^{22}$~cm$^{-2}$. 
Average observed fluxes are in units of 
10$^{-11}$~erg~cm$^{-2}$~s$^{-1}$. X--ray luminosities are in units of 10$^{33}$~erg~s$^{-1}$ 
in the 2--10 keV energy band 
and have been calculated adopting distances determined by \citet{Rahoui2008} from optical spectroscopy 
of the supergiant companions 
(4.9~kpc for IGR~J16479$-$4514, 2.7~kpc for XTE~J1739$-$302
and 3.6~kpc for IGR~J17544$-$2619). 
IGR~J18410$-$0535 is located in the direction of the
Scutum Arm, at a distance probably comprised between 1 and 10 kpc \citep{Bamba2001}; 
for this source we will arbitrarily assume 5~kpc.\label{sfxts:tab:spectra}
} 
\smallskip
\smallskip
\begin{tabular}[c]{ccccccc}
\hline\noalign{\smallskip}
\hline\noalign{\smallskip}
Source  &  $N_{\rm H}$                       &  parameter     &       Av. Observed Flux     &   Av. Luminosity  & $\chi^{2}_{\rm red}$ (dof) \cr
\noalign{\smallskip\hrule\smallskip}
Absorbed power law  	     &               &   $\Gamma$     &       (2--10 keV)  &     (2--10 keV)              &            \cr
\noalign{\smallskip\hrule\smallskip}
   IGR~J16479$-$4514         &  $7.7^{+1.0}_{-0.9}$    &$1.6_{-0.2}^{+0.2}$           &  2.03  & 87  &   0.969 (112)   \cr
     XTE~J1739$-$302         &  $3.3^{+0.04}_{-0.04}$  &$1.4_{-0.4}^{+0.4}$           &  0.37  & 3.9  &  1.16 (21)    \cr
   IGR~J17544$-$2619         &  $3.2^{+1.2}_{-0.9}$    &$2.1^{+0.6}_{-0.5}$           &  0.32  & 6.3  &  0.916 (16)   \cr
   IGR~J18410$-$0535         &  $4.2^{+1.7}_{-1.1}$    &$1.6^{+0.6}_{-0.4}$           &  0.35  & 13  &  0.59 (20)    \cr
\noalign{\smallskip\hrule\smallskip}
Absorbed blackbody  	     &              &  $kT_{\rm bb}$   &      \cr
\noalign{\smallskip\hrule\smallskip}
   IGR~J16479$-$4514         &  $4.5^{+0.6}_{-0.5}$    &$1.6\pm{0.1}$  & 1.85 &  62  &  0.954 (112)   \cr
     XTE~J1739$-$302         &  $1.6^{+0.6}_{-0.5}$    &$1.5\pm{0.2}$  & 0.32 &  3.0 &  1.08 (21)      \cr
   IGR~J17544$-$2619         &  $1.5^{+0.6}_{-0.4}$    &$1.1\pm{0.1}$  & 0.26 &  4.4 &  1.109 (16)     \cr
   IGR~J18410$-$0535         &  $2.0^{+0.9}_{-0.6}$    &$1.5\pm{0.2}$  & 0.30 &  9.6 &  0.72 (20)      \cr
\noalign{\smallskip\hrule\smallskip}
\end{tabular}
\end{table*}

\end{document}